\documentclass[showpacs,twocolumn,superscriptaddress,prl]{revtex4}%
\usepackage{amsmath}
\usepackage{graphicx}
\usepackage{bm}
\usepackage{amsfonts}
\usepackage{amssymb}%
\begin{document}
\title{Detection of current-induced spins by ferromagnetic contacts}
\author{\.{I}. Adagideli}
\affiliation{Department of Physics and Astronomy, University of British Columbia, 6224
Agricultural Road, Vancouver, B.C. V6T 1Z1, Canada}
\altaffiliation{current address: \textit{Institut f{\"u}r Theoretische Physik, Universit{\"a}t
Regensburg, D-93040, Germany} }
\author{G.E.W. Bauer}
\affiliation{Kavli Institute of Nanoscience, TU Delft, Lorentzweg 1, 2628 CJ Delft, The Netherlands}
\author{B.I. Halperin}
\affiliation{Lyman Laboratory, Department of Physics, Harvard University, Cambridge, MA
02138, USA}
\date{\today}

\begin{abstract}
Detection of current-induced spin accumulation via ferromagnetic contacts is
discussed. Onsager's relations forbid that in a two-probe configuration spins
excited by currents in time-reversal symmetric systems can be detected by
switching the magnetization of a ferromangetic detector contact. Nevertheless,
current-induced spins can be transferred as a torque to a contact
magnetization and affect the charge currents in many-terminal configurations.
We demonstrate the general concepts by solving the microscopic transport
equations for the diffuse Rashba system with magnetic contacts.

\end{abstract}

\maketitle

The notion that netto spin distributions can be generated by electric currents
in the bulk of non-magnetic semiconductors with intrinsic spin-orbit (SO)
interaction has been predicted~\cite{Levitov,Edelstein} and experimentally
confirmed~\cite{Katoaccum}. The related spin Hall effect, causing
accumulations of spins at the edges, has also been observed \cite{exp}. It can
be extrinsic, \textit{i.e.} caused by impurities with SO
scattering~\cite{DP,extrinsic} or intrinsic due to an SO split band
structure~\cite{Murakami,Sinova}. However, all experiments to date detected
the current-induced spins optically \cite{metal}. An important remaining
challenge for theory and experiment is to find ways to transform the novel
spin accumulation (SA) and spin currents (SCs) into voltage differences and
charge currents in micro- or nanoelectronic circuits in order to fulfill the
promises of spintronics.

In this Letter, we address the possibility to generate a spin-related signal
to be picked up by ferromagnetic contacts. This signal can be in the form of a
voltage change or a torque acting on the magnetization of the ferromagnet (FM). In
practice, this raises technical difficulties due the conductance mismatch
\cite{Schmidt} that can be solved~\cite{Crowell} and are not addressed here.
We rather focus on conceptual problems that are related to the voltage and
torque signals generated by current-induced spins~\cite{vanWees}.
The Onsager relations for the conductance~\cite{Onsager} forbid voltage based
detection of current-induced spins by a ferromagnetic lead in a two-probe
setup within linear response. We sketch a microscopic picture of the
physics and formulate a semiclassical scheme in the diffuse limit. We address
spin detection in a multi probe geometry, calculate the Hall conductivity and
compare it to the anomalous Hall effect. We also discuss the possibility to
construct spin filters not based on FMs, but on conductors with
spatially modulated SO interactions.

\begin{figure}[ptb]
\includegraphics[width=8cm]{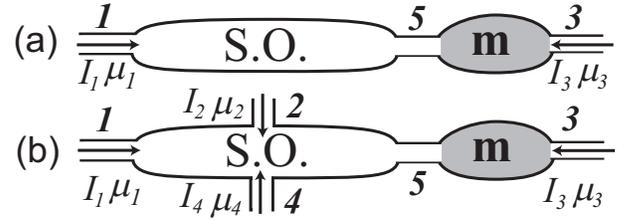}
\caption{Current-Voltage configuration for (a)~two- and (b)~four-probe
setup.}%
\label{FIG:cur_setup}%
\end{figure}We first address the symmetry properties in terms of the Onsager
relations for the (linear) conductance
\cite{Onsager,REF:Casimir,REF:Buttiker86,REF:Hankiewicz}. A generic
SO-Hamiltonian $H$ involves products of velocity and spin operators that are
invariant under time reversal. Even when spin-degenerate bands are split due
to a broken inversion symmetry, the Kramers degeneracy remains intact. The
Hamiltonian of the combined SO$|$F system has the symmetry $TH(\mathbf{m}%
)T^{-1}=H(-\mathbf{m})$, where $\mathbf{m}$ is a unit vector in the direction
of the magnetization of the FM and $T$ the time-reversal operator. We
now focus on a general multi-probe setup (see Fig.~\ref{FIG:cur_setup}b for a
4-probe setup). The currents in the leads and the respective chemical
potentials of the reservoirs are related in linear response as $I_{i}=\sum
_{j}G_{ij}\mu_{j}$. The elements of the conductance matrix can then be
expressed as a commutator of the respective current operators $\hat{I}_{i}(t)$
and $\hat{I}_{j}(t)$ in the leads $i$ and $j$:
$ G_{ij}=\lim_{\omega\rightarrow0}\mathrm{Re}\,\int_{0}^{\infty}dt
(e^{i\omega t}/\omega)
\langle\lbrack\hat{I}_{i}(t),\hat{I}_{j}(0)]\rangle,
$
where the brackets $\langle\cdots\rangle$ denote either ground state
expectation value or thermal averaging. For reversed magnetization we can
relate the states that enter the averaging with those of the original
magnetization using
$T$. By inserting $T^{-1}T$ between the
operators in the Kubo formula it follows $G_{ij}(-\mathbf{m})=G_{ji}%
(\mathbf{m})$. For the two-probe configuration (Fig.~\ref{FIG:cur_setup}a), the currents in the two leads
are equal and opposite, by current conservation, and there is only one
conductance $G$. This implies $G(\mathbf{m})=G(-\mathbf{m})$. Thus, it is
not possible to detect the SA by reversing the magnetization or current
direction in this case.

Voltage signals can be measured in a three or more terminal setup, however. We
focus on the
current/voltage
configuration in Fig.~\ref{FIG:cur_setup}b with
$I_{1}=-I_{3}$, $I_{2}=-I_{4}$, $eV_{1}=\mu_{3}-\mu_{1}$ and $eV_{2}=\mu
_{4}-\mu_{2}$~\cite{REF:Casimir} appropriate, e.g., to a set up where leads 2
and 4 are connected together by a resistor, while leads 1 and 3 are connected
by a battery. Then $I_{i}=\sum_{j=1,2} (-1)^{i+j}\alpha_{ij}V_{j}$, where the
coefficients $\alpha_{ij}$ can be found in Eqs. (4.a-4d) of Ref.
\cite{REF:Buttiker86}. The Onsager relations can then be expressed as:
$\alpha_{ij}(\mathbf{m})=\alpha_{ji}(-\mathbf{m})$\label{EQ:Onsager}. If we
choose (say) $I_{1}=0$, the relation between the applied current and the
spin-Hall voltage is: $I_{2}=V_{1}(\alpha_{11}\alpha_{22}-\alpha_{12}%
\alpha_{21})/{\alpha_{12}}$. Upon switching the magnetization direction we
obtain the same formula except for the change of $\alpha_{21}$ to $\alpha
_{12}$ and vice versa: $G_{H}(\mathbf{m})=\frac{\alpha_{21}}{\alpha_{12}}%
G_{H}(-\mathbf{m}). $ Since in general $\alpha_{21}\neq\alpha_{12},$ $G_{H}$
changes as
$\mathbf{m}$
is switched. Our analysis implies the
equivalence of two Hall measurements: (i)~setting $I_{1}$ to zero and
detecting $V_{1}$ generated by an applied $I_{2}$ and (ii)~setting $I_{2}$
equal to zero and detecting $V_{2}$ \cite{Valenzuela}. In other words, driving
a current $I_{2}$ through the system and detecting the spin Hall voltage with
a ferromagnetic contact is equivalent to driving a SA into the SO region via a
FM that leads to a real Hall voltage detectable by normal
contacts. We note that the symmetries of the joint SO$|$F region is
identical to a bulk FM with SO interaction. A
SC
cannot be detected by merely switching $\mathbf{m}$ or
current direction in a two-probe setup~\cite{Ref:AHE}.

Further insight to the Onsager relations can be obtained by considering the
microscopic scattering processes at the SO$\vert$F interface. For simplicity,
we take the FM to be halfmetallic and with a matched electronic
structure, \textit{i.e}., electrons with spin parallel (antiparallel) to
$\mathbf{m}$ are transmitted\ (reflected). The conductance is proportional to
the sum of transmission probabilities including multiple scattering processes
at the SO$|$F interface and impurities. When the spin polarized by the point
contact is a majority electron in the FM, it is directly transmitted
with probability unity. When we reverse the magnetization direction, the
electron is initially completely reflected. However, the reflected electron
subsequently scatters at impurities, experiencing the effective Zeeman field
due to the SO coupling that causes the spin to precess. It eventually
approaches the interface again with a different spin orientation. Only the
component parallel to the majority spin direction is transmitted. For the
reflected antiparallel component the game starts all over again. By repeated
scattering at the interface, the electron is eventually transmitted for both
magnetization directions with equal probability. Note the similarity of this
picture with the reflectionless tunneling at a superconducting
interface~\cite{REF:reflesstunn}. An electron reflected once at a
ferromagnetic Hall contact, on the other hand, has a finite probability to
escape into the drain, thus leaving a voltage signal of its spin. When the SC
is generated by a point contact and injected into a ballistic normal
conductor~\cite{Eto} the argument is clearer: although an up-spin
electron leaving the point contact is reflected by an attached FM
with antiparallel magnetization, the subsequent spin-flip reflection at the
point contact leads to a transmission probability that is the same for a
parallel magnetization, in agreement with the Onsager relations.

We now focus on a microscopic transport theory for a simple model Hamiltonian,
the Rashba spin-split two-dimensional electron gas (R2DEG) with SO coupling
$H^{R}=(\alpha/\hbar)\mathbf{p}\cdot({\bm\sigma}\times\hat{\bm z})$, where
$\mathbf{p}$ is the momentum operator, $\{\sigma_{i}\}$ are the Pauli spin
matrices, and $\alpha$ parameterizes the strength of the SO interaction. For a
setup as in Fig.~\ref{FIG:cur_setup} and in the diffuse limit with
spin-independent s-wave scatterers we compute the voltage signal in the
ferromagnetic lead.
The diffusion equation for the (momentum-integrated)
density matrix $\rho(E)=n(E)+\mathbf{s}(E)\boldsymbol{\cdot\sigma}$ has been obtained
by Ref.s~\cite{Mishchenko,Burkov}. We use the notation of Ref.~\cite{inanc} for
diffusion equations (Eq.s~(6-8) in Ref.~\cite{inanc}) and for the expression of SC in terms
of $\rho(E)$ (Eq.(9) in Ref.~\cite{inanc}).
We
define $L_{s}=\sqrt{D\tau_{s}}$, $\lambda=K_{s-c}/D$ and
$\eta=K_{p}L_{s}/D$.
These diffusion equations are the leading terms in a
gradient expansion, and require for their validity that the SA gradient is
small compared to the transport mean free path $v_{F}\tau$. Generally, this
requires $L_{s}>v_{F}\tau$ and results strictly hold only in the dirty limit,
$\xi\ll1$. Nevertheless, since results are well behaved for all $\xi$ they
might be useful beyond the regime of formal validity. We consider a weak
FM with diffusion equation $\nabla^{2}s_{m}=s_{m}/L_{sF}^{2}$ for the
SA component $s_{m}$ parallel to the magnetization, where $L_{sF}$ is the
spin-flip relaxation length in the FM. The charge and SCs in the
FM read $\mathbf{j}(E)=-\nu_{F}D_{F}(\boldsymbol{\nabla}n+\delta
D\boldsymbol{\nabla}s_{m})$ and $\mathbf{j}^{m}(E)=-\nu_{F}D_{F}%
\big(\boldsymbol{\nabla}s_{m}+(\delta D/4)\boldsymbol{\nabla}n\big)$ where
$\nu_{F}D_{F}=(\nu_{+}D_{+}+\nu_{-}D_{-})/2$, $\delta D=(\nu_{+}D_{+}-\nu
_{-}D_{-})/(\nu_{F}D_{F})$, $\nu_{\pm}$ and $D_{\pm}$ are the density of
states and diffusion constants of the majority and minority spin electrons.

We chose a simple model for the matching conditions between a R2DEG and a
FM: the charge current as well as the SC polarized in the
magnetization direction of the FM are conserved at the interface,
whereas the SC polarized perpendicularly to the magnetization direction
transfers a torque onto the magnetization by being absorbed at the
interface~\cite{Slonczewski,Nunez}. We have a now complete set of
matching/boundary conditions at the R2DEG$\mid$F interface: $m_{i}%
(\bm{n}\cdot\bm{j}_{R}^{i})=m_{i}(\bm{n}\cdot\bm{j}_{F}^{i}),$ $\bm{m}\times
\bm{s}_{R}=0$ and $\bm{m}\cdot\bm{s}_{R}=\bm{m}\cdot\bm{s}_{F}$, where the
subscripts F and R refer to the FM and the R2DEG respectively. The
first condition holds when the interface does not cause additional spin
relaxation. The second one requires that the perpendicular SC is dephased in
the FM on a length scale that is much smaller than the mean free
path. The third condition,~\textit{i.e}. continuity of the SA in the
magnetization direction, has been useful for purposes of
illustration before~\cite{old,Kovalev}, but does not hold for general
interfaces~\cite{Galitski}.

\begin{figure}[ptb]
\mbox{ \includegraphics[width=\columnwidth]{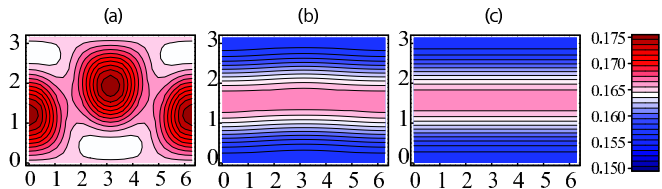}}
%
\caption{The potential drop in units of $J_{p}\delta D^{2}L_{s}/4D\nu_{R}$ as
a function of the magnetization direction for $\Gamma(1-\delta D^{2}/4)=5$ and
$\xi=0.2$ (a), $\xi=1$ (b) and $\xi=5$ (c). The $y$-axis is the azimuthal
angle $\theta$ and $x$-axis is the polar angle $\phi$ of $\mathbf{m}$.}%
\label{FIG:Cond}%
\end{figure}
Solving the diffusion equations with boundary conditions is simplified by
decomposing
the SC across the interface
into an injection
current from the FM and an out-diffusion current of the SA in the
R2DEG. The former is a linear function of $\mathbf{m}$ while the latter
depends on the bulk SA direction. We define two vector functions
$\boldsymbol{\mu}$ and $\mathbf{f}$ at point $\mathbf{r}$ at the
interface as the $\mathbf{m}$-dependent and independent part of the
normal derivative of $\mathbf{s}$:
\begin{equation}
L_{s}(\mathbf{n}\cdot\boldsymbol{\nabla})s_{i}|_{r}=\big(\mu_{i}%
(\mathbf{m})|\mathbf{s}|+\lambda L_{s}^{2}(\mathbf{z}\times\mathbf{J}_{p}%
)_{i}f_{i}/D\nu_{R}\big)\big\vert_{r}\nonumber
\end{equation}
Here $\mu_{i}$ is a linear function of $\mathbf{m}$, $\mathbf{n}$ is the unit
normal vector at the interface, and $\mathbf{J}_{p}$ is the local current density.

For a (wide contact) two-terminal configuration $\mathbf{n}$, $\mathbf{J}_{p}%
$, $\boldsymbol{\mu}$ and $\mathbf{f}$ are constant over the interface. We
choose the $x$-direction to be the current direction and $y$-direction
parallel to the interface. We then obtain the SA at the interface to order
$(\alpha/k_{F})^{2}$ as:
\begin{align}
\mathbf{s}|_{0} & =\mathbf{m}J_{p}(\delta D/4-m_{y}\lambda L_{s}%
)/\Lambda\left(  \mathbf{m}\right),\\
\Lambda\left(  \mathbf{m}\right)&
=L_{s}^{-1}D\nu_{R}\mathbf{m}%
\cdot\boldsymbol{\mu}+L_{sF}^{-1}D_{F}\nu_{F}(1-\delta D^{2}/4).
\end{align}
Here, $\mu_{x}=(\sqrt{2}-1)(\kappa_{+}+\kappa_{-})m_{x}-(2-\sqrt{2})2\eta
m_{z}$, $\mu_{y}=m_{y}$ and $\mu_{z}=2\eta(\sqrt{2}-1)m_{x}+(2-\sqrt
{2})(\kappa_{+}+\kappa_{-})m_{z}$, where $2\kappa_{\pm}^{2}=3-4\eta^{2}%
\pm(1-24\eta^{2}+16\eta^{4})^{1/2}$ and $\mathrm{Re}\kappa_{\pm}>0$. Note that
these expressions do not depend on $f_{i}$. In contrast to the SA at the
interface the potential drop
$
\phi_{FR}=-J_{p}\delta D^{2}/8\Lambda
$
is invariant under $\mathbf{m}\rightarrow-\mathbf{m}$, consistent with the
Onsager relations (Fig.~\ref{FIG:Cond}). In the clean limit, the terms that do
not involve $m_{z}$ agree with previous results in the absence of SO
interaction~\cite{old}. The anomalous $m_{z}$ dependence is due to a different
spin relaxation length for the $z$-component of the spin. A similar effect has
been observed in numerical simulations~\cite{Pareek}. Although $\phi_{FR}$
does not depend on the direction of $\mathbf{m}$ in the $xy$-plane, all
higher order corrections in $\alpha/k_{F}$ have an even angular dependence.
The resistance modulation differs from the tunneling magnetoresistance found
by R\"{u}ster \textit{et al}.~\cite{Ruester} in that here the SO interaction
is in the normal metal rather than in the FM.

As shown above, the spin Hall voltage does not need to be invariant under
reversal of the magnetization of the FM. Using the same diffusion
equations, we calculate the extra potential drop $\phi_{H}$ at a
ferromagnetic \textquotedblleft Hall\textquotedblright\ lead. Now the
interface of the contact lies parallel to the (bulk) current direction $x$.
When the charge current through the ferromagnetic lead is biased to zero, we
obtain
\[
\phi_{H}=\frac{J_{p}^{\prime}\delta D}{2\Lambda}\left(  \frac{\alpha
\xi\mathbf{m}\cdot\mathbf{f}}{v_{F}\sqrt{1+4\xi^{2}}}-\frac{m_{z}\xi^{2}}%
{4\pi\hbar D\nu_{R}(1+4\xi^{2})}\right) ,
\]
that changes sign with the magnetization direction. Here $J_{p}^{\prime}$ is
the local average current density parallel to the Hall lead, $\mu_{x}=m_{x}$, $\mu
_{y}=(\sqrt{2}-1)(\kappa_{+}+\kappa_{-})m_{y}-(2-\sqrt{2})2\eta m_{z}$,
$\mu_{z}=2\eta(\sqrt{2}-1)m_{y}+(2-\sqrt{2})(\kappa_{+}+\kappa_{-})m_{z}$,
$f_{x}=0$, $f_{y}=(\kappa_{+}+\kappa_{-})(\sqrt{2}-1)$, and $f_{z}=(\sqrt
{2}-1)2\eta$.

\begin{figure}[ptb]
\mbox{ \includegraphics[width=8.5cm]{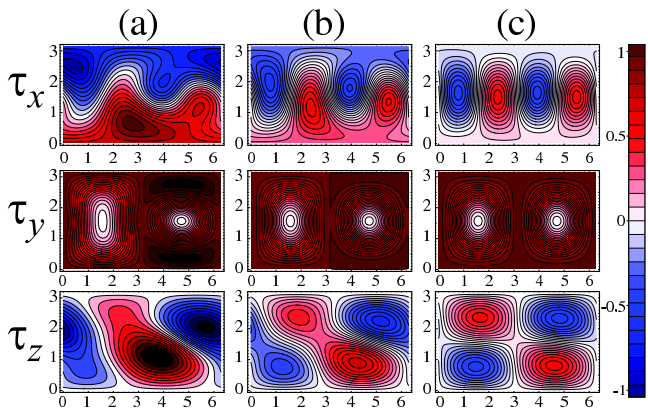}} \caption{Components of the
spin transfer torque, scaled by a factor $J_{p}\lambda L_{s}$, plotted for the
set of parameters $\xi=1$, $\zeta=5$ and (a)~$\Gamma=1$, (b)$\Gamma=5$ and
(c)~$\Gamma=25$. The $y$-axis is the azimuthal angle $\theta$ and $x$-axis is
the polar angle $\phi$ of $\mathbf{m}$.}%
\label{FIG:torque}%
\end{figure}

We now focus on the transverse component of the angular momentum that is
transferred to the FM as a torque on the
magnetization~\cite{Slonczewski,Nunez}, which is not restricted to certain
symmetries under
reversal of $\mathbf{m}$
by the Onsager relations. For the
two-probe setup, we obtain:
\begin{align}
\tau_{i}  &  =J_{p}\lambda L_{s}\Bigg[\delta_{i2}-m_{y}m_{i}+\frac{D_{R}%
\nu_{R}}{L_{s}\Lambda}\left(  \frac{\delta D}{4\lambda L_{s}}-m_{y}\right)
\nonumber\\
&  \times\left(  \mu_{i}-m_{i}(\mathbf{m}\cdot\boldsymbol{\mu})-\eta
^{1/3}(\delta_{i3}m_{x}-\delta_{i1}m_{z})\right)  \Bigg]
\label{EQ:torque}
\end{align}
We notice both even and odd contributions under magnetization reversal. Their
ratio is controlled by the dimensionless parameter $\zeta=\delta D/4\lambda
L_{s}$ and the parameter related to the conductance mismatch $\Gamma=\nu
_{F}D_{F}L_{s}/\nu_{R}D_{R}L_{sF}$, both
typically much larger
than unity. For $\zeta\ll\Gamma$ $\left(  \zeta\gg\Gamma\right)  $ the torque
is an even (odd) function of $m_{i}$. For general parameters the torque
displays
a surprising complexity (Fig.~\ref{FIG:torque}). In
the limit $\zeta\ll\Gamma$,
the even terms dominate and it is
possible to switch the magnetization by a charge current. The switching
dynamics will be complicated by the odd terms if $\zeta\lesssim\Gamma$. These
effects are
rotating $\mathbf{m}$ around the accumulation direction (which
is due to spin precession in the SO region and is due to the term
proportional to $\eta^{1/3}$ in the expression above), and
pulling $\mathbf{m}$ out of the plane depending
on whether $m_{z}$ is greater
than or less than zero.

Assuming a nm scale thick ferromagnetic film (such as Fe)
on InAs in the clean limit, we estimate, using Slonczewski's
expressions~\cite{Slonczewski}, the critical current density for
magnetization-switching to be of the order of $10^{0}\mathrm{A/cm}$
(equivalent to a bulk current density of $10^{6}\mathrm{A/cm^{2}}$).
This result will increase by a factor $g_{S}/g$, where $g_{S}$ and
$g$ are the Sharvin and total spin conductance of the contact
respectively. It is possible to reduce critical currents strongly by
reducing the magnetization of FM. This can be achieved by
choosing a ferromagnetic semiconductor (that should be
\textit{n}-type in the case or InAs \cite{Kio}) that is furthermore
close to its Curie temperature~\cite{Chiba}. Excessive heating can be
avoided by short-time pulses \cite{Chiba}. Small-angle magnetization
dynamics is induced at much smaller currents and can be detected optically for e.g.
a three-terminal geometry.
Finally, we
mention that dominant SO coupling in 2D heavy hole gas has a
$k^{3}$ dependence which does not generate any bulk
SA~\cite{SAhole}. Therefore SA effects predicted
here
will be relatively weak for most 2DHGs. The
$k$-linear SO coupling can be increased in 2DHGs by applying
strain.

We finally note that spin filters can be fabricated without FMs by
spatial modulation of the SO
coupling.
This can be useful for the R2DEG in
which $\alpha$ can be modulated by an electric field. In a symmetric but
externally gated quantum well, the inversion-symmetry-breaking electric field,
and thus the sign of $\alpha$ can be reversed. The region
in which $\alpha$ is reversible would be able to detect the SC
generated in
another one, since the
resistance of the $\alpha|\alpha$ junction
differs from that of the $\alpha|-\alpha$ junction.

In conclusion, we addressed current-induced spin related voltage and torque signals in
ferromagnetic contacts. We presented the conductance and spin transfer torque
due to the spin-polarized currents generated in a spin-orbit coupled region in
two- and many-terminal devices and discussed their symmetry properties in
terms of the Onsager relations. Symmetry prohibits the detection of spins in
two-probe conductance measurement via a magnetization switch. In contrast, it
is possible to detect a torque signal, distinctive of SCs, even in a
two probe setting. For the four-probe Hall type measurement, both current
induced SA and the spin Hall current may be detected by voltages, where the
contribution of the spin Hall current to the Hall conductivity bears close
similarity to the anomalous Hall effect.

B.I.H. would like to thank H.-A. Engel and E. G. Mishchenko for useful
discussions. This work was supported by the FOM, EU Commission FP6 NMP-3
project 05587-1 \textquotedblleft SFINX\textquotedblright, NSF grants
PHY-0117795 and DMR-0541988 and NSERC discovery grant number R8000.


\begin{thebibliography}{99}                                                                                               %


\bibitem {Levitov}F. T. Vas'ko and N. A. Prima, Sov. Phys. Solid State
\textbf{21}, 994 (1979);
L.S. Levitov~\textit{et al}.,
Zh. Eksp. Teor. Fiz. \textbf{88}, 229 (1985).

\bibitem {Edelstein}V. M. Edelstein, Sol. Stat. Commun. \textbf{73}, 233
(1990);
J.I. Inoue~\textit{et al}.,
Phys. Rev. B \textbf{67},\ 033104 (2003).

\bibitem {Katoaccum}
Y.K. Kato~\textit{et al}.,
Nature \textbf{427}, 50 (2004);
Y.K. Kato~\textit{et al}.,
Phys. Rev. Lett. \textbf{93}, 176601 (2004); In hole systems: A. Yu. Silov
\textit{et al}., Appl. Phys. Lett. 85, 5929 (2004); S. D. Ganichev \textit{et
al}., cond-mat/0403641 (unpublished).

\bibitem {exp}
Y.K. Kato~\textit{et al}.,
Science \textbf{306}, 1910 (2004);
J. Wunderlich~\textit{et al}.,
Phys. Rev. Lett. \textbf{94}, 047204 (2005);
V. Sih~\textit{et al}., Nature Phys. 1, 31-35 (2005).

\bibitem {metal}Several very recent experiments report electric signals of the
extrinsic spin Hall effect in metals : E. Saitoh \textit{et al}., Appl. Phys.
Lett. 88, 182509 (2006); S.O. Valenzuela and M. Tinkham, Nature \textbf{442},
176--179 (2006); T. Kimura \textit{et al.}, cond-mat/0609304.

\bibitem {DP}M.I.~Dyakonov and V.I.~Perel, JETP Lett.~\textbf{33},
467 (1971).

\bibitem {extrinsic}J. E. Hirsch, Phys. Rev. Lett. \textbf{83}, 1834 (1999);
S. Zhang, Phys. Rev. Lett. \textbf{85}, 393 (2000); R.V. Shchelushkin and A.
Brataas, Phys. Rev. B \textbf{71}, 045123 (2005); J.~Hu \textit{et al}., Int.
J. Mod. Phys. B \textbf{17}, 5991 (2003) ; S.-Q.~Shen, Phys. Rev. B
\textbf{70}, 081311(R) (2004); D.~Culcer \textit{et al}., Phys. Rev. Lett.
\textbf{93}, 046602 (2004); N.A.~Sinitsyn \textit{et al}., Phys. Rev. B
\textbf{70}, 081312 (2004); A.A.~Burkov \textit{et al}., Phys. Rev. B
\textbf{70}, 155308 (2004).

\bibitem {Murakami}
S.~Murakami~\textit{et al}.,
Science \textbf{301}, 1348 (2003); Phys. Rev. B 69, 235206 (2004).

\bibitem {Sinova}
J.~Sinova~\textit{et al}., Phys. Rev. Lett. \textbf{92}, 126603 (2004).

\bibitem {Schmidt}
G. Schmidt~\textit{et al}., Phys. Rev. B \textbf{62}, R4790 (2000).

\bibitem {Crowell}
X. Lou~\textit{et al}., cond-mat/0602096; P. Crowell \textit{c.s.},
unpublished; P. Cheng \textit{et al}., cond-mat/0608453

\bibitem {vanWees}B. J. van Wees, Phys. Rev. Lett. 84, 5023 (2000);
P. R. Hammar \textit{et al}.,
Phys. Rev. Lett. 84, 5024 (2000)

\bibitem {Onsager}L. Onsager, Phys. Rev. B 38, 2265 (1931)

\bibitem {REF:Casimir}H. B. G. Casimir, Rev. Mod. Phys. \textbf{17}, 343 (1945).

\bibitem {REF:Buttiker86}M. B\"{u}ttiker, Phys. Rev. Lett. \textbf{57}, 1761 (1986).

\bibitem {REF:Hankiewicz}Onsager-like relations have been discussed, assuming
an inversion symmetric setup, by E. M. Hankiewicz \textit{et al}., Phys. Rev.
B 72, 155305 (2005).

\bibitem {Valenzuela}The spin-Hall signal in metallic devices \cite{metal} has
been measured by configuration (ii).

\bibitem {Ref:AHE}J. Inoue and H. Ohno, Science \textbf{309}, 2004 (2005).

\bibitem {REF:reflesstunn}
I. K. Marmorkos~\textit{et al}.,
Phys. Rev. B \textbf{48}, 2811 (1993);
C. W. J. Beenakker~\textit{et al}.,
Phys. Rev. Lett. \textbf{72}, 2470 (1994);
B. J. van Wees~\textit{et al}., Phys. Rev. Lett. \textbf{69}, 510 (1992);

\bibitem {Eto}
M. Eto~\textit{et al}.,
J. Phys. Soc. Jap.
\textbf{74}, 1934 (2005); P.G. Silvestrov and E.G. Mishchenko,
Phys. Rev. B \textbf{74}, 165301 (2006)

\bibitem {Slonczewski}J.C. Slonczewski, J. Magn. Magn. Mater. \textbf{159}, L1 (1996).

\bibitem {Nunez}A. S. N\'{u}\~{n}ez and A.H. MacDonald, Sol. State Commun.
\textbf{139}, 31 (2006).

\bibitem {Mishchenko}
E.G.~Mishchenko~\textit{et al}.,
Phys. Rev. Lett. \textbf{93}, 226602 (2004).

\bibitem {Burkov}
A.A.~Burkov~\textit{et al}.,
Phys.Rev. B \textbf{70}, 155308 (2004).

\bibitem {inanc}\.{I}. Adagideli and G.E.W. Bauer, Phys. Rev. Lett.
\textbf{95}, 256602 (2005).

\bibitem {Kovalev}
A. A. Kovalev~\textit{et al}.,
Phys. Rev. B \textbf{66}, 224424 (2002).

\bibitem {Galitski}
V. M. Galitski~\textit{et al}.,
cond-mat/0601677.

\bibitem {old}M. Johnson and R. H. Silsbee, Phys. Rev. Lett. \textbf{55}, 1790
(1985);
P. C. van Son~\textit{et al}.,
Phys. Rev. Lett. \textbf{58}, 2271 (1987).

\bibitem {Pareek}T.P. Pareek, Phys. Rev. B \textbf{70}, 033310 (2004)

\bibitem {Ruester}C. R\"{u}ster et al., Phys. Rev. Lett. \textbf{94}, 027203 (2005)

\bibitem {Inoue}
J.I.~Inoue~\textit{et al}.,
Phys. Rev. B \textbf{70}, 041303(R) (2004).

\bibitem {pdoped}S. Murakami, Phys. Rev. B \textbf{69}, 241202(R) (2004).

\bibitem {Kio}G. Kioseoglou et al., Nature Materials \textbf{3}, 799 (2004).

\bibitem {Chiba}
D. Chiba \textit{et al}., Phys. Rev. Lett., \textbf{93}, 216602 (2004).

\bibitem {SAhole}
A. V. Shytov~\textit{et al}.,
Phys. Rev. B \textbf{73}, 075316 (2006);
T. L. Hughes~\textit{et al}.,
cond-mat/0601353 ;
O. Bleibaum, S. Wachsmuth, cond-mat/0602517.
\end{thebibliography}
\end{document}